\documentclass[aps,prc,twocolumn,groupedaddress,showpacs,showkeys,nofootinbib,fleqn]{revtex4}

\usepackage{epsfig,amsmath,amssymb,graphicx,tabularx,dcolumn,bm,longtable} 

\newcommand{\beqy}{\begin{eqnarray}}
\newcommand{\eeqy}{\end{eqnarray}}
\newcommand{\bmlet}{\begin{subequations}}
\newcommand{\emlet}{\end{subequations}}
\newcounter{saveeqn}

\def\gsimeq{\,\,\raise0.14em\hbox{$>$}\kern-0.76em\lower0.28em\hbox {$\sim$}\,\,}
\def\lsimeq{\,\,\raise0.14em\hbox{$<$}\kern-0.76em\lower0.28em\hbox {$\sim$}\,\,}
\def\beqy{\begin{eqnarray}}
\def\eeqy{\end{eqnarray}}
\def\<{\langle}
\def\>{\rangle}
\def\bmlet{\begin{mathletters}}
\def\emlet{\end{mathletters}}

\newcommand{\ud}{\mathrm{d}}

\newcommand{\kBCS}{|\text{HFB}\>}
\newcommand{\bMF}{\< \text{MF}|}
\newcommand{\kMF}{|\text{MF}\>}

\begin{document}

\title{Further explorations of Skyrme-Hartree-Fock-Bogoliubov mass formulas.\\ 
       III: Role of particle-number projection}

\author{M. Samyn}
\email{msamyn@astro.ulb.ac.be}
\affiliation{Institut d'Astronomie et d'Astrophysique, 
             ULB - CP226, 1050 Brussels, Belgium}

\author{S. Goriely}
\affiliation{Institut d'Astronomie et d'Astrophysique, 
             ULB - CP226, 1050 Brussels, Belgium}

\author{M. Bender}
\affiliation{Service de Physique Nucl\'eaire Th\'eorique et de Physique Math\'ematique,
             Universit\'e Libre de Bruxelles - CP229, 1050 Brussels, Belgium}

\author{J.M. Pearson}
\affiliation{D\'ept. de Physique, Universit\'e de Montr\'eal, 
             Montr\'eal (Qu\'ebec), H3C 3J7 Canada}

\date{August 28 2004}

\begin{abstract} Starting from HFB-6, we have constructed a new mass table, referred to
as HFB-8,  including all the 9200 nuclei lying between the two drip lines over the
range of $Z$ and
$N \ge 8$ and $Z \le 120$. It differs from HFB-6 in that the wave function is projected
on the exact particle number. Like HFB-6, the isoscalar effective mass
$M_s^*$ is constrained to the value $0.80M$ and the pairing is density independent. The
rms errors of the mass-data fit is 0.635~MeV, i.e. better
than almost all our previous  HFB mass formulas. The extrapolations of this new mass
formula out to the drip lines do not differ significantly from the previous
HFB-6 mass formula. 
\end{abstract}



\maketitle

\section{Introduction}
\label{into}

In the last few years we have been able to construct complete mass tables 
by the Hartree-Fock-Bogoliubov (HFB) method \cite{sghpt02,gshpt02,sg03,gs03}, with 
the parameters of the underlying forces being fitted to essentially all of
the available mass data. This paper is the third in a series of studies of 
possible refinements and modifications to our HFB-2 mass formula \cite{gshpt02}, 
the first of our models that was able to give a satisfactory fit to the new 
data that had accumulated since the 1995 Atomic Mass Evaluation \cite{aw95}.
The most obvious reason for making such changes would be to improve the data 
fit, but there is also a considerable astrophysical interest in being able to 
generate different mass formulas even if no significant improvement in the data
fit is obtained. 
The main point here is that the r-process of stellar 
nucleosynthesis proceeds through the formation of nuclei that are so highly 
neutron rich that their properties cannot be measured but must be inferred by
extrapolating the properties of known nuclei, and there is no guarantee that
mass formulas giving equivalent mass-data fits will still give the same masses 
out towards the neutron drip line. Moreover, even if they do, it is still 
possible that the underlying model (forces) will give different results for 
other properties relevant to the r-process (see Section 1 of the first paper of
this series \cite{sg03}, referred to here as paper I). Predictions for the
equation of state of neutron star matter could likewise differ.

In paper I \cite{sg03} we discussed the role of a possible density dependence 
of the pairing interaction, while in paper II \cite{gs03} we examined the question
of the effective nucleon mass to be imposed on the Skyrme force. In the present 
paper we extend our HFB model by the restoration of the particle-number symmetry.
In Section \ref{model} we will recall the main features of the earlier HFB models 
relevant for the present study. In Sect.~\ref{sect2} we present our formalism 
for particle-number restoration. In Section \ref{sect3} we describe
a new fit of the force parameters to the mass data, with the resulting 
parameter set being labelled BSk8, and discuss its ability to predict 
masses and radii. A complete new mass table, HFB-8, is constructed
and compared with the earlier HFB-6 mass table in Sect.~\ref{sect4}.
Conclusions are drawn in Sect.~\ref{concl}.

%********************************************************************************************

\section{Summary of the model}
\label{model}
For convenience we recall some of the essential features of these HFB 
models. They are are based on conventional Skyrme forces of the form  
\begin{eqnarray}
\label{1}
v_{ij}^{ph} & = & t_0(1+x_0P_\sigma)\delta({{\bf r}_{ij}})\nonumber\\ & + &
t_1(1+x_1P_\sigma)\frac{1}{2\hbar^2}\{p_{ij}^2\delta({{\bf r}_{ij}}) +h.c.\}\nonumber\\
& + & t_2(1+x_2P_\sigma)\frac{1}{\hbar^2}{\bf p}_{ij}.\delta({\bf r}_{ij})
 {\bf p}_{ij}\nonumber\\ & + & \frac{1}{6}t_3(1+x_3P_\sigma)\rho^\gamma\delta({\bf
r}_{ij})\nonumber\\ & + & \frac{i}{\hbar^2}W_0(\mbox{\boldmath$\sigma_i+\sigma_j$})
           {\bf .p}_{ij}\times\delta({\bf r}_{ij}){\bf p}_{ij}  
\end{eqnarray} 
in the particle-hole (ph) channel, and in the particle-particle (pp) channel a
$\delta$-function pairing force acting between like nucleons treated in the full
Bogoliubov framework
\begin{equation}
\label{2}
v_{ij}^{pp}= V_{\pi q}~\left[1-\eta
\left(\frac{\rho}{\rho_0}\right)^\alpha\right]~
\delta(\mbox{\boldmath$r$}_{ij}) \ ,
\end{equation} 
where $\rho \equiv \rho({\bf r})$ is the local density, and $\rho_0$ is its
equilibrium value in symmetric infinite nuclear matter (INM). 
Actually, it was only with models HFB-3 \cite{sg03}, HFB-5 \cite{gs03}, 
and HFB-7 \cite{gs03} that the possibility of a density dependence in the
pairing force was admitted; in all our other HFB models, HFB-1 \cite{sghpt02}, 
HFB-2 \cite{gshpt02}, HFB-4 \cite{gs03}, and HFB-6 \cite{gs03} we had $\eta$ = 0, 
as will be the case in the present paper.  

An important aspect relating to $\delta$-function pairing 
forces concerns the cutoff to be applied to the space of single-particle 
(s.p.) states over which the force is allowed to act: both BCS and Bogoliubov 
calculations diverge if this space is not truncated \cite{b02,by02}. However, making 
such a cutoff is not simply a computational device but is rather a vital
part of the physics, pairing being essentially a finite-range phenomenon.
To represent such an interaction by a
$\delta$-function force is thus legitimate only to the extent that all
high-lying excitations are suppressed, although how exactly the truncation of
the pairing space should be made will depend on the precise nature of the real,
finite-range pairing force. It was precisely our ignorance on this latter point
that allowed us in Ref.\ \cite{gshpt02} to exploit the cutoff as a new degree of 
freedom: we found there an optimal mass fit with the spectrum of s.p. states 
$\varepsilon_i$ confined to lie in the range
\begin{equation}
\label{3}
E_F - \varepsilon_{\Lambda} 
\le \varepsilon_i 
\le E_F + \varepsilon_{\Lambda}
,
\end{equation}
where $E_F$ is the Fermi energy of the nucleus in question, and
$\varepsilon_{\Lambda}$ is a free parameter. We shall adopt the same 
parametrization in the present paper.

The mean-field states for even and odd particle number have different
structure. While they are HFB states for even particle number, they are
one-quasiparticle excitations for odd particle number
\begin{eqnarray}
\kMF & = & \kBCS 
           \qquad \text{even particle number,} \\
\kMF & = & \hat{a}^\dagger_b \kBCS 
           \qquad \text{odd  particle number.}
\end{eqnarray}
Throughout this paper, we use the canonical basis to represent the 
HFB states.
The blocking of the unpaired particle is not calculated completely 
self-consistently. Doing so requires breaking the time-reversal and axial 
symmetries, and adding extra fields to the Hamiltonian. 
This is still too costly for a large-scale mass fit as performed here. Therefore
we use an approximation, where the pairing correlations in odd nuclei 
are determined by removing the last nucleon \cite{sghpt02}.
The occupation probability 
of the last occupied level is later corrected by adding the occupation 
contribution of the unpaired nucleon. The last nucleon therefore does 
not contribute to the pairing (it does through the pairing tensor 
in the spherical configuration), and the corresponding level should 
be treated accordingly. For the unprojected mean-field states, 
the occupation probability of the last occupied level $k$ is 
\beqy
\label{w_k}
w_k^2=\frac{(d(k)-2)v_k^2+1}{d(k)} \ ,
\eeqy
where the degeneracy $d(k)$ equals 2 for deformed nuclei and $2j+1$
for spherical ones ($j$ being the angular momentum quantum number);
$v_k^2$ is the occupation probability obtained from the HFB 
equations for the level $k$ without the blocked nucleon.
To compensate for the approximations done for odd particle number, we
incorporate what is missing into the effective interaction and shall allow 
the pairing-strength parameter $V_{\pi q}$ to be different for 
an odd particle number ($V_{{\pi q}}^-$) than for an even particle number 
($V_{{\pi q}}^+$). For example, the pairing force between neutrons
depends on whether $N$ is even or odd. 
We also allow the pairing-strength $V_{\pi q}$ to be different for protons 
and neutrons, as in all our previous mass models.
Another feature of the HFB-2 model \cite{gshpt02} that we retain here is to add
to the energy calculated by the HFB model a phenomenological Wigner term of the 
form
\begin{eqnarray}
E_W &=& V_W\exp\Bigg\{-\lambda\Bigg(\frac{N-Z}{A}\Bigg)^2\Bigg\} \nonumber\\
& + & V_W^{\prime}|N-Z|\exp\Bigg\{-\Bigg(\frac{A}{A_0}\Bigg)^2\Bigg\} ~. 
\label{4}
\end{eqnarray}
The method used to solve the HFB equations for Skyrme interactions has been presented 
earlier in Ref.~\cite{sghpt02}. 
%
%********************************************************************************************
%
\section{Restoration of the particle-number symmetry}
\label{sect2}
Mean-field approaches, such as the HFB used here, establish an intrinsic frame of the
nucleus and consequently break several symmetries of the Hamiltonian and the wave
function in the laboratory frame \cite{rs80,bh03}. In particular, finite nuclei
break translational invariance, deformed nuclei rotational invariance,
reflection asymmetric shapes the parity symmetry, and the HFB framework the
particle-number symmetry. These symmetry breakings are required to include the desired
correlations to the modeling (as multi-particle-multi-hole states), but at the same time
gives rise to an admixture of excited states to the calculated ground state. The broken
symmetries can be restored rigorously by projecting the wave function on the exact
quantum numbers. A simpler procedure aims at estimating the contribution  to the binding
energy in a suitable approximation, and to add the resulting  correction to the binding
energy. We adopted such a procedure in some of our previous mass formulas, in particular
to estimate the center-of-mass (cm) correction from the recoil energy, and the rotational
correction within the cranking model \cite{tgpo00}.  
As far as the cm correction is concerned, the approximate prescription of Butler {\it et al.}
\cite{bs84} was replaced in \cite{gs03} by a more fundamental calculation of the the recoil 
energy. 
However, so far no attempt to restore the particle-number symmetry has been undertaken
in any of our HFB calculations related to global mass fit. The present paper is devoted
to the projection of the wave function on the exact number of particles after a variation
that includes the approximate Lipkin-Nogami projection before variation (referred to
as PLN), and its impact on the mass fit. 

The pair correlations included in the mean-field HFB wave function $\kMF$ are known to 
break the particle-number symmetry, as an HFB state is not an eigenstate of the
particle-number operator. While the expectation value of the particle number
$\bMF \hat{N} \kMF$ is constrained to have the desired number $N$ of 
nucleons on the average via a Lagrange multiplier (the chemical potential), 
its dispersion
\beqy
\label{disper}
\bMF (\Delta\hat{N})^2 \kMF
& = & \bMF \hat{N}^2 \kMF - \bMF \hat{N} \kMF^2
      \nonumber \\
& = & 2\sum_{k \lessgtr 0} u_k^2 v_k^2
,
\eeqy 
never vanishes in the presence of pairing. We use the standard
notation, where $v_k^2=1-u_k^2$ is the occupation probability of the level $k$.

The particle-number symmetry is restored by applying a 
particle-number projection operator $\hat{P}^N$ on the mean-field state
\cite{rs80,hb93,ae01,ae02} 
\beqy
|\Phi\>
= \frac{\hat{P}^N \kMF}
       {\bMF \hat{P}^N \kMF^{1/2}}
\eeqy
which eliminates all components of the HFB state with a particle number 
different from $N$. $|\Phi\>$ is a normalized eigenstate of the
particle-number operator. The particle-number projector is given by
\beqy
\label{phin}
\hat{P}^N
= \frac{1}{2\pi}\int_0^{2\pi}d\phi\ \, e^{i\phi(\hat{N}-N)}
.
\eeqy
The integration interval of the gauge angle $\phi$ can be reduced to $[0,\pi]$ for
intrinsic wave functions with a well-defined ``number parity'' quantum number, i.e., \ for
even-even systems as  well as odd systems if treated in the blocking approximation (see
below). For practical applications, the integral over the gauge angle 
in the particle-number projection operator has to be discretized,
\beqy
\label{discr} 
P_n(M_g)=\frac{1}{M_g}\sum_{m=1}^{M_g} e^{i\phi_m(\hat{N}-n)}
.
\eeqy 
We use the Fomenko prescription \cite{fo70}, 
where $\phi_m=\pi (m-1)/M_g$ and $M_g$ is the number of 
angles in the interval $[0,\pi]$ used in the calculation.
It can be shown that this particular choice of the projector
removes all unwanted contaminations of the states up to 
$n \pm 2M_g$ \cite{rs80}. 
In order to avoid certain numerical problems that arise at 
$\phi=\frac{\pi}{2}$ when $v^2_k$ has accidentally the value $0.5$ \cite{ae01}
we restrict ourselves to odd values of $M_g$, choosing finally $M_g = 11$ 
for both protons and neutrons in all calculations.
The particle number dispersion is subsequently checked
to verify that the projection actually worked.

The particle-number symmetry restoration is performed in two steps. 
First, an approximate projection before variation is performed within the Lipkin-Nogami
prescription \cite{N64,PN73,rn96,fo97,ae01}, where we take only the pairing part 
of the interaction into account when calculating $\lambda_2$. In a second step, the
converged HFB states are then exactly projected as described by Eq.~(\ref{phin}).
As usually done, we constrain the HFB states to the same particle number that we 
project out exactly afterwards. In such a projection after variation approach, 
the main task for the Lipkin-Nogami scheme is to enforce the presence of pair correlations 
in the HFB states for all $N$ and $Z$ and all deformations to avoid the breakdown
of pairing for small level densities. 
This two-step approach was shown to be an excellent approximation to the exact 
self-consistent particle-number projection \emph{before} variation \cite{ae02}.

The nuclear mass and all other observables are calculated from the projected state
applying a generalized Wick's theorem. The basic contractions needed for that are 
given by \cite{ae01}
\beqy
\rho_{kk'}(\phi)&\equiv&
 \frac{\bMF a^{\dag}_k a_{k'} e^{i\phi \hat{N}} \kMF}
      {\bMF \text{MF} (\phi)\>}
\nonumber \\
  &=&\frac{v^2_k
e^{2i\phi}}{u^2_k+v^2_ke^{2i\phi}}\delta_{kk'}=\rho_{\bar{k}\bar{k}'}(\phi) \
,\label{rhodef} \\
\kappa_{k\bar{k}'}(\phi)&\equiv&
 \frac{\bMF a^{\dag}_k a^{\dag}_{\bar{k}'} e^{i\phi \hat{N}} \kMF}
      {\bMF \text{MF} (\phi)\>} \nonumber \\
  &=&\frac{-u_kv_k}{u^2_k+v^2_ke^{2i\phi}}\delta_{kk'}=-\kappa_{\bar{k}k'}(\phi) \ ,
\label{kappadef} \\
\tilde{\kappa}_{k\bar{k}'}(\phi)&\equiv&
 \frac{\bMF a_k a_{\bar{k}'} e^{i\phi \hat{N}} \kMF}
      {\bMF \text{MF} (\phi)\>}
\nonumber \\
  &=&\frac{u_k v_k
e^{2i\phi}}{u^2_k+v^2_ke^{2i\phi}}\delta_{kk'}=-\tilde{\kappa}_{\bar{k}k'}(\phi)
\label{kappatdef}
,
\eeqy 
where we introduced the shorthand notation 
$| \text{MF} (\phi)\> \equiv e^{i\phi \hat{N}} \kMF$ for a MF state rotated 
in gauge space. We employ the phase convention 
($u_{\bar{k}},v_{\bar{k}})=(u_k,-v_k)$, where $k$ denotes
single-particle states with positive angular momentum projection on the symmetry 
axis, and $\bar{k}$ the corresponding time-reversed states with negative angular 
momentum projection on the symmetry axis. 
For the blocked orbital in a mean-field state with odd particle number, the
contractions of the level $k$ are determined by removing the blocked nucleon,
since it should not enter the particle number projection. Of course, the
contribution of this unpaired nucleon is added to Eq.~(\ref{rhodef})
afterwards (see, for instance, Eq.~(\ref{rrkk})).
In order to express the expectation value of observables in the projected state, 
it is useful to introduce normalized $\phi$-dependent overlaps
\beqy 
O(\phi,n)  \equiv  
\frac{\bMF \text{MF} (\phi) \> \, e^{-in\phi}}
     {\sum_{m=1}^{M_g} \bMF \text{MF} (\phi_m)\> \, e^{-in\phi_m}}\ .
\eeqy 
Based on Eq.~(\ref{rhodef}), the number of particles of the projected state
can then be written as
\beqy
\label{Neq}
\< \Phi | \hat{N} | \Phi \>
= \sum_{m=1}^{M_g} O(\phi_m,n) \; {\rm Tr}[\rho(\phi_m)]\ ,
\eeqy
which reduces to the familiar  $\<\hat{N}\>=2\sum_{k>0}v_k^2$ if $M_g=1$.

The particle-number projected energy depends on the $\phi$-dependent 
generalized densities and is calculated by an integration over the spatial 
coordinates ${\bf r}$ and the summation over the two sets of gauge angles 
$\phi^q_m$ ($q=n,p$). The total mean-field energy can therefore be expressed  as
\beqy 
E_{\text{mf}}
&=& \sum_{m_n=1}^{M_g}\sum_{m_p=1}^{M_g} 
    O (\phi^n_{m_n},n) \, O(\phi^p_{m_p},p)
    \nonumber \\
& & \times \int \ud{\bf r} \; {\cal E}_{\text{mf}}({\bf r},\phi^n_{m_n},\phi^p_{m_p})
,
\eeqy 
where ${\cal E}_{\text{mf}}$ is the Skyrme energy density functional (obtained
with Eq.~(3a) of \cite{tgpo00} and the $\phi-$dependent densities).
The pairing energy is calculated similarly:
\beqy 
E_{\text{pair},q}&=& 
\sum_{m_n=1}^{M_g} \sum_{m_p=1}^{M_g}
O(\phi^n_{m_n},n) \, O(\phi^p_{m_p},p) \nonumber \\
 && \times \int \ud{\bf r} \; {\cal E}_{\text{pair},q}({\bf r},\phi^n_{m_n},\phi^p_{m_p})
,
\eeqy 
where
\beqy 
{\cal E}_{\text{pair},q}({\bf r},\phi^q_{m_q},\phi^q_{m_q})&=& 
V_{\pi q}
{\rm Tr}[\kappa(\phi^q)]
{\rm Tr}[\tilde{\kappa}(\phi^q)] \nonumber \\ &&
\hspace*{-2cm}
\times \left( 1 - \eta\bigg[\frac{(\rho({\bf r},\phi^n_{m_n},\phi^p_{m_p})}
                                 {\rho_0}       \bigg]^{\alpha} \right)
.
\eeqy 
For the density-dependence in $E_{\text{mf}}$ and $E_{\text{pair}}$ we choose
the mixed density obtained from Eq.~(\ref{rhodef}), as done in most applications 
involving the mixing of different mean-field states. This choice, however, is 
not unique, see the discusion in Refs.~\cite{ae01,ro02} and references given therein.
%
%********************************************************************************************
%
\subsubsection{Decomposition of mean-field states}
The weight of the $n'$-nucleon wave function in the decomposition of the 
unprojected $n$-particle HFB state $| \text{MF}_n \>$ in given by
the overlap 
\beqy
\label{Overlap} 
{\cal O}_{n'} &=& \< \text{MF}_n | \hat{P}^{n'} | \text{MF}_n \> 
.
\eeqy 
This quantity can be used to illustrate the particle number 
dispersion of the HFB states.

\begin{figure}[t!]
\begin{center}
\centerline{\epsfig{figure=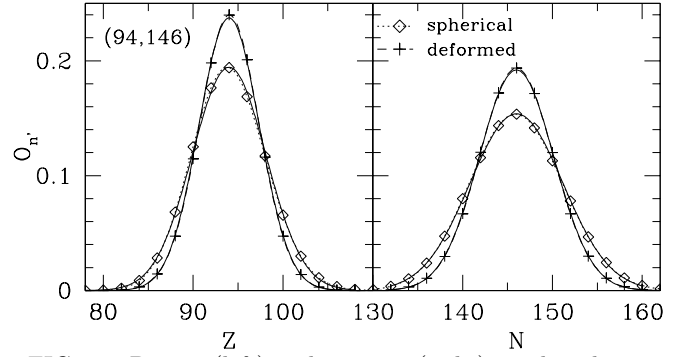,scale=.45}}
\vspace*{-1.2cm}
\end{center}
\caption{\label{pu240} Proton (left) and neutron (right) overlaps between the
unprojected HFB wave function of $^{240}$Pu and its state projected on $n'$
($N$ or $Z$) nucleons; diamonds refer to spherical overlaps, crosses to deformed 
ones; the full line corresponds to a Gaussian centered on $n=94$ (protons) or 
146 (neutrons) and of standard deviation $\sigma^2$ equal to the particle-number 
dispersion.}
\end{figure}

An example is given in Fig.~\ref{pu240} for $^{240}$Pu, where we compare the
spherical configuration with the deformed equilibrium one, which has a
quadrupole moment of $Q=2782$~fm$^2$. The decomposition is significantly
different. In lowest order, the distribution of the weights has a Gaussian 
shape \cite{fo97}
\beqy
\label{OverlapGauss}
{\cal O}_{n'} \simeq \frac{2}{\sqrt{2\pi \sigma^2}}
                \exp\Big (-\frac{(\bMF\hat{N}\kMF-n')^2}{2 \sigma^2}\Big )
,
\eeqy
where the width of the Gaussian $\sigma^2 = \bMF (\Delta \hat{N})^2 \kMF$ is 
determined by the particle-number dispersion of the unprojected state, 
Eq.~(\ref{disper}). In the specific case of $^{240}$Pu, 
the Gaussian is more spread for the spherical shape, where pairing is 
stronger due to the larger level density around the Fermi surface.
%
%--------
%
%
%--------
%

\begin{figure}[t!]
\begin{center}
\centerline{\epsfig{figure=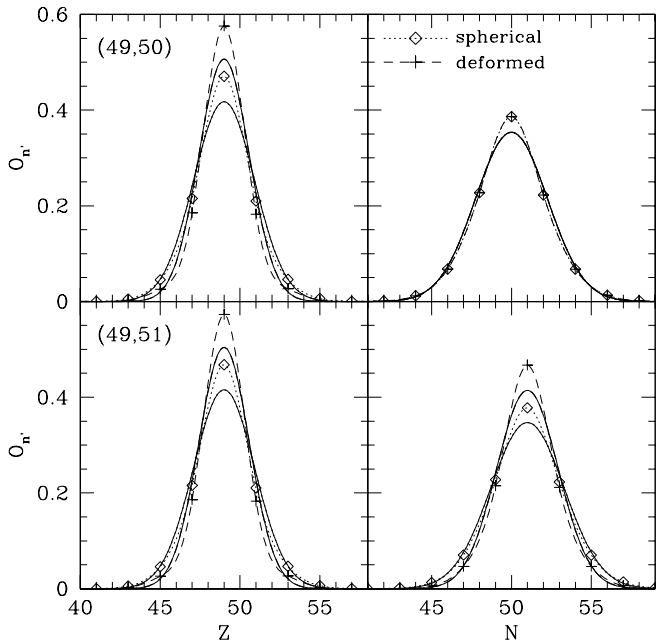,scale=.45}}
\vspace*{-1.2cm}
\end{center} 
\caption{\label{in100} Same as Fig.~\ref{pu240} for the spherical 
$^{99}$In and $^{100}$In istopes.
The Gaussian approximation, Eq.~\ref{OverlapGauss} (thin line with dots or dashes 
obtained from the spherical or deformed code) cannot describe 
the pronounced shell closure. For $Z=49$ or $N=51$, the overlaps
obtained with the deformed code differ significantly from the overlaps 
calculated assuming spherical symmetry, due to a different implementation 
of the blocking approximation.  }
\end{figure}

The different blocking prescription in spherical and deformed configurations 
leads to significant 
variations in the weight of neighbouring nuclei in the wave function, 
as seen in Fig.~\ref{in100} for the spherical $^{99}$In and $^{100}$In isotopes:
the blocked nucleon in the spherical case is not removed from a 
doubly-degenerate level (to leave it empty) but
removed from the $g_{9/2}$ level, which is equivalent to the removal 
of 20$\%$ of a nucleon on the last five doubly-degenerate levels of a 
deformed configuration. The weight of the wave functions of adjacent nuclei 
is therefore higher (in the $Z=49$ or $N=51$ In isotopes) if
the spherical code is used. It is also observed that the deviation 
between Eq.~(\ref{OverlapGauss}) and Eq.~(\ref{Overlap}) is more pronounced, 
due to a shell closure effect.

For consistency, the cm correction should be calculated from the particle-number 
projected state. Written concisely, the gauge angle-dependent cm energy
of the particle-number projected state reads
\beqy 
{\mathcal E}_{cm}(\phi)&=&-\frac{\hbar^2}{2M}\Big [
2\sum_{k>0}\rho_{kk}\Delta_{kk} 
 +2\sum_{k,l>0}(\rho_{kk}\rho_{ll} +
\kappa_{k\bar{k}}\tilde{\kappa}_{l\bar{l}})\nonumber \\
 && \qquad (\mbox{\boldmath$\nabla$}_{kl}      \cdot \mbox{\boldmath$\nabla$}^*_{kl}
 +\mbox{\boldmath$\nabla$}_{k\bar{l}}\cdot \mbox{\boldmath$\nabla$}^*_{k\bar{l}}) \Big ]
.
\eeqy 
For odd nuclei, $(\rho_{kk}\rho_{ll} + \kappa_{k\bar{k}}\tilde{\kappa}_{l\bar{l}})$ 
becomes 
\beqy\label{rrkk}
\lefteqn{\Big [\frac{(d(k)-2\delta_b(k))\rho_{kk}+\delta_b(k)}{d(k)}\ } \nonumber \\
&& \frac{(d(l)-2\delta_b(l))\rho_{ll}+\delta_b(l)}{d(l)}
+ \kappa_{k\bar{k}}\tilde{\kappa}_{l\bar{l}}\Big ] \quad ,
\eeqy
with $\delta_b(k)=1$ if the blocked nucleon is on the level $k$ (0 otherwize).
The cm energy (for one nucleon species $q$) is then 
\beqy 
E_{cm,q} = \sum_{m=1}^{M_g} O(\phi_m,q) \, {\mathcal E}_{cm,q}({\bf r},\phi_m)
.
\eeqy
Finally, note that as far as the rotational correction is concerned, we will restrict
ourselves to the simplified cranking formula, although ideally the rotational symmetry
should also be restored with projection techniques. So far, in all our global mass fits, 
we adopted the rotational correction calculated from the HFB states
\beqy\label{erot} 
E_{\text{rot}} = \frac{\bMF \hat{J}^2 \kMF} {2{\mathcal I}} \ ,
\eeqy 
where $\hat{J}$ is the total angular momentum operator in the intrinsic frame 
and ${\mathcal I}$ the moment of inertia.
The cranking model \cite{in54,in56,be61} gives this latter
quantity as
\beqy
\label{Icr}
{\mathcal I}_{cr}&=&2\sum_{k,k'>0}\frac{|\<k|J_x|k'\>|^2}{E_k+E_{k'}}
(u_kv_{k'}-u_{k'}v_{k})^2
,
\eeqy
where the summation runs over quasi-particle s.p. states, the
$E_k$ are the corresponding quasi-particle energies, the matrix elements are
calculated in the canonical basis, and $v_k^2 = 1 - u_k^2$ are the
corresponding occupation probabilities.

While the cranking model is in general quite satisfactory, a major problem
in our mass-model applications arises from the fact that {\it a priori} we
do not know which nuclei will be spherical (or quasi-spherical), for which nuclei
a numerical problem arises from the 0/0 indeterminacy in Eq. (\ref{erot});
$E_{rot}$ must vanish, of course, for spherical nuclei. A way around this
problem, introduced in Ref.~\cite{pa91}, is to take for the
${\mathcal I}$ the linear combination
\beqy\label{Imixte}
{\mathcal I}_{mix}=a{\mathcal I}_{cr}+(1-a){\mathcal I}_{rig} \quad ,
\eeqy
where ${\mathcal I}_{rig}$ is the rigid-body moment of inertia
\beqy\label{7}
{\mathcal I}_{rig}&=&\frac{M}{3}(2 R^2 A+\frac{1}{2}Q) \quad ,
\eeqy
in which $R$ is the rms matter radius of the nucleus, $M$ the nucleon mass, and
$Q$ the quadrupole moment. The value of $a$ was originally taken to be 0.8,
but was reduced to 0.75 in Ref. \cite{gtp01} and kept unchanged in all subsequent mass
formulas (HFB-1--HFB-7). However, we see from Table~\ref{bsk8_expi} that with this
prescription the deformation energies ${\Delta}E_{iso}$ of highly deformed shape
isomers, i.e., their energy relative to the ground state of the nucleus in question,
are badly underestimated, and can even be negative. 
For this reason we adopt for the rotational energy the phenomenological prescription
\beqy
\label{Erotcrth}
E_{rot}=b E_{rot}^{crank}~\tanh(c|\beta_2|) \ ,
\eeqy
or equivalently introduce a new prescription based on the cranking
value of the moment of inertia modified as follows:
\beqy
\label{Icrth}
{\mathcal I}_{crth}&=&\frac{1}{b}{\mathcal I}_{cr}\coth(c|\beta_2|)
. 
\eeqy
Here, the dimensionless quadrupole moment $\beta_2$ is defined 
as a  function of the quadrupole moment $Q_2$ and the reduced radius $R_0$ by
$\beta_2=\sqrt{5\pi}Q_2/(3AR_0^2)$. Experimental information on the mass of well
deformed nuclei, but also the energy of the shape isomers in the actinide
region is used to determine the free parameters $(b,c)$.
Values of $(b,c)$ range from (0.6,5.5) to (0.7,3.5). For HFB-8, we adopt $b=0.65$ and
$c=4.5$.  We find that this prescription not only
is equivalent to the ``mixed" prescription (\ref{Imixte}) for ground states, i.e., for
masses, and likewise avoids any problems in the spherical limit, but also leads to
considerable improvements in the estimates for the shape isomers, as can be seen from
the penultimate column of Table~\ref{bsk8_expi}.

\begingroup
\squeezetable
\begin{table}
\caption{\label{bsk8_expi} Comparison between experimental
\cite{bl80,bf88,jaeri,hg01} and theoretical (reflection symmetric assumed) energies of
shape isomers obtained with the HFB+PLN(BSk8) model using three
different rotational correction prescriptions: Eq. (\ref{Icrth}) (modified
cranking), Eq. (\ref{Imixte}) (mix with $a=0.87$), and Eq. (\ref{7}) (rigid).
The last two lines show the rms and mean deviations, respectively, compared to
experiment. All quantities in MeV.}
\begin{ruledtabular}
\begin{tabular}{ccccccc}
\hline\noalign{\smallskip}
   Z &   N &   A &  exp & crth &   mix & rig \\
\noalign{\smallskip}\hline\noalign{\smallskip}
  90 & 140 & 230 &  2.25 & 2.6 &   1.6 &  2.2 \\
     & 141 & 231 &  2.3  & 2.4 &   1.3&  2.1 \\
     & 143 & 233 &  2.3  & 2.5 &   1.5&  2.3 \\
  92 & 143 & 235 &  2.5  & 2.3 &   1.2 &  1.9 \\
     & 144 & 236 &  2.3  & 2.3 &   1.2 &  1.8 \\
     & 145 & 237 &  2.2  & 2.2 &   0.9 &  1.9\\
     & 146 & 238 &  2.6  & 2.4 &   1.4 &  2.0\\
     & 147 & 239 &  1.9  & 1.9 &   0.6&  1.5 \\
  93 & 144 & 237 &  2.7  & 1.7 &   0.7  &  1.3\\
     & 145 & 238 &  2.3  & 1.6 &   0.6 &  1.4\\
  94 & 141 & 235 &  2.6  & 1.5 &   0.4  &  1.1\\
     & 143 & 237 &  2.3  & 1.9 &   0.8 &  1.4 \\
     & 144 & 238 &  2.4  & 1.8 &   0.7 &  1.3\\
     & 145 & 239 &  2.2  & 1.8 &   0.6 &  1.3\\
     & 146 & 240 &  2.25 & 1.9 &   0.8  &  1.5\\
     & 147 & 241 &  1.9  & 1.3 &   0.1 &  0.9\\
     & 149 & 243 &  1.7  & 2.0 &   0.8  &  1.6\\
     & 150 & 244 &  2.0  & 2.2 &   1.0 &  1.7\\
  95 & 144 & 239 &  2.4  & 1.4 &   0.4  &  1.0\\
     & 145 & 240 &  2.6  & 1.4 &   0.4  &  1.1\\
     & 146 & 241 &  2.2  & 1.7 &   0.6  &  1.3\\
     & 147 & 242 &  2.3  & 1.0 &  -0.2 &  0.6 \\
     & 148 & 243 &  2.0  & 1.7 &   0.6 &  1.3\\
     & 149 & 244 &  1.6  & 1.6 &   0.5 &  1.3\\
  96 & 145 & 241 &  2.0  & 1.1 &   0.0  &  0.7\\
     & 146 & 242 &  1.8  & 1.3 &   0.2 &  0.8\\
     & 147 & 243 &  1.5  & 1.1 &  -0.2  &  0.6\\
     & 148 & 244 &  1.04 & 1.5 &   0.3 &  1.0 \\
     & 149 & 245 &  1.7  & 1.3 &  0.0 &  0.9\\
  97 & 147 & 244 &  2.0  & 0.6 &  -1.0  &  0.1\\
\noalign{\smallskip}\hline\noalign{\smallskip}
$\sigma$   &  &  &       & 0.64 & 1.62 & 0.95\\
$\epsilon$ &  &  &       & 0.40 & 1.54 & 0.80\\
\noalign{\smallskip}\hline
\end{tabular}
\end{ruledtabular}
\end{table}
\endgroup

The rotational correction should be accompanied by corrections for vibrational 
zero-point motion. To the best of our knowledge, there exists no strategy to
estimate them properly and at the same time include them in a global mass fit as ours 
with current computing resources. 
Common strategies are to calculate RPA correlations \cite{st02,ba04}, 
usually restricted to spherical nuclei, or to use the generator 
coordinate method \cite{bbh03}. More research in this direction is necessary 
in the future.

%%%%%%%%%%%%%%%%%%%%%%%%%%%%%%%%%%%%%%%%%%%%%%%%%%%%%%%%%%%%%%%%%%%%%%%%%%%%%%%%
%
\section{The mass fit}
\label{sect3}
\begin{figure}[ht]
\centerline{\epsfig{figure=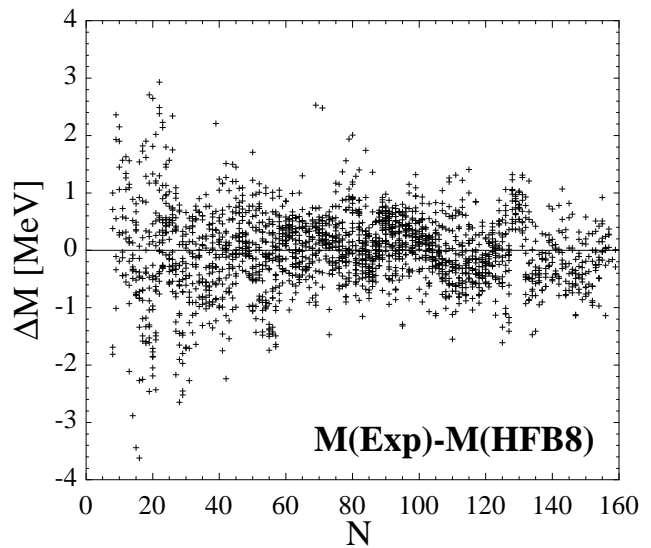,scale=.37}}
\caption{\label{bsk8_exp} differences between experimental and calculated mass excesses
as a function of the neutron number $N$ for the HFB-8 mass tables} 
\end{figure}

To study the impact of the PLN framework, we start from the BSk6 force
\cite{gs03}, keeping some of its characteristics: an isoscalar effective mass 
$M^*_s/M$ constrained to 0.8, the symmetry energy $J$ to 28~MeV, the $\gamma$ 
exponent in the $t_3$ term of Eq.(~\ref{1}) to 1/4, and a density-independent 
pairing interaction (i.e., $\eta=0$). 
The results of the HFB-8 mass fit of the new HFB+PLN model are presented in 
Table ~\ref{tab_rms_8}, where for comparison we also show the results of the 
HFB-6 model. Since the latter was fitted to the 2135 nuclei with  
$Z,N \geq 8$ whose masses had been measured and compiled
in the unpublished  2001 Atomic Mass Evaluation (AME) of Audi and Wapstra 
\cite{aw01} we show results for both this data set and the 2149 measured masses
of the updated AME that was released and published at the end of 2003 
\cite{aw03}. Experimental and calculated mass excesses are compared 
in Fig.~\ref{bsk8_exp}. 
\begin{table}[ht]
\centering
\caption{Rms ($\sigma$) and mean ($\bar{\epsilon}$) errors (in MeV) in the predictions
of masses $M$ obtained with the BSk6 and BSk8 forces with respect to the 2135 nuclei in
the 2001 AME \cite{aw01} and the 2149 nuclei in the 2003 AME \cite{aw03}. 
The last two lines correspond to the rms and mean  errors (in fm) in the
predictions of the 523 measured charge radii ($r_c$).} 
\label{tab_rms_8}
 \vspace{.5cm}
 \tabcolsep=.3cm
 \begin{ruledtabular}
 \begin{tabular}{ccc}
\hline\noalign{\smallskip}
  & BSk6 & BSk8  \\
\noalign{\smallskip}\hline\noalign{\smallskip}
 $\sigma(M)$              (2135 nuclei) & 0.684 & 0.659   \\
 $\bar{\epsilon}(M)$      (2135 nuclei) & 0.021 & 0.005   \\
 $\sigma(M)$              (2149 nuclei) & 0.666 & 0.635   \\
 $\bar{\epsilon}(M)$      (2149 nuclei) & 0.014 & 0.009   \\
 $\sigma(r_c)$            (523 nuc)     & 0.0262 & 0.0250 \\
 $\bar{\epsilon}(r_c)$   (523 nuc)     & -0.0028 & 0.0047 \\
\noalign{\smallskip}\hline
\end{tabular}
\end{ruledtabular}
\end{table}

The parameters of the force BSk6 and BSk8 are compared in Table~\ref{tab_sky8},
while the macroscopic parameters, i.e., the parameters relating to (semi-) infinite
nuclear matter calculated for these forces, are
shown in Table~\ref{tab_par8}. These include, in addition to the previously defined
parameters, $a_v$, the energy per nucleon at equilibrium in symmetric infinite nuclear
matter, $M^*_v/M$ the ratio of the isovector
effective nucleon mass at density $\rho_0$ to the real nucleon mass $M$, 
$K_v$, the incompressibility, $G_0$ and $G_0^{\prime}$ the Landau parameters 
as defined in Ref.~\cite{gs81}, $\rho_{frmg}$, the density at which 
neutron matter flips over into a ferromagnetic state that
has no energy minimum and would collapse indefinitely \cite{kw94}, $a_{sf}$,
the surface coefficient, and $Q$, the surface-stiffness coefficient
\cite{ms69} (the meaning of $G_0$, $G_0^{\prime}$, and $\rho_{frmg}$ is critically
discussed in \cite{bh03,bd02}).

\begin{table}[ht]
 \centering
 \caption{Skyrme-force and pairing-force parameters of BSk6 and BSk8}
\label{tab_sky8}
 \vspace{.2cm}
 \tabcolsep=.2cm
 \begin{ruledtabular}
 \begin{tabular}{ccc}
\hline\noalign{\smallskip}
  & BSk6 & BSk8 \\
\noalign{\smallskip}\hline\noalign{\smallskip}
  $t_0$ {\scriptsize [MeV fm$^3$]}   & -2043.317 & -2035.525  \\
  $t_1$ {\scriptsize [MeV fm$^5$]}   & 382.127    & 398.8208    \\
  $t_2$ {\scriptsize [MeV fm$^5$]}   & -173.879   & -196.0032   \\
  $t_3$ {\scriptsize [MeV fm$^{3+3\gamma}$]} & 12511.7 & 12433.36   \\
  $x_0$                              &  0.735859  &  0.773828   \\
  $x_1$                              &  -0.799153 & -0.822006   \\
  $x_2$                              & -0.358983  & -0.389640   \\
  $x_3$                              & 1.234779   &  1.309331   \\
  $W_0$ {\scriptsize [MeV fm$^5$]}   & 142.4     & 147.8    \\
  $\gamma$                           &  1/4       & 1/4   \\
  $V^+_n$ {\scriptsize [MeV fm$^3$]} &  -321.2    & -314.0  \\
  $V^-_n$ {\scriptsize [MeV fm$^3$]} &  -337.9    & -329.8  \\
  $V^+_p$ {\scriptsize [MeV fm$^3$]} &  -324.5    & -293.0  \\
  $V^-_p$ {\scriptsize [MeV fm$^3$]} &  -342.4    & -309.9  \\
  $\eta$                             & 0          &  0  \\
  $\alpha$                           & 0          &  0  \\
  $\varepsilon_{\Lambda}$ {\scriptsize [MeV]}  & 17   &  17  \\ 
  $V_W$ {\scriptsize [MeV]}          & 1.76 & 1.85  \\
  $\lambda$                          & 700  & 780   \\
  $V_W^{\prime}$ {\scriptsize [MeV]} & 0.58 & 0.66  \\
  $A_0$                              & 28   & 26    \\
  $a$ (Eq.(\ref{Imixte}))            & 0.75 & -  \\
  $b,c$ (Eq.(\ref{Erotcrth}))        &   -  & 0.65,4.5 \\
 \noalign{\smallskip}\hline
 \end{tabular}
 \end{ruledtabular}
 \end{table}
% \vspace{-1cm}
%
 \begin{table}[ht]
 \centering
 \caption{Macroscopic parameters of the forces BSk6 and BSk8}
\label{tab_par8}
 \vspace{.2cm}
 \tabcolsep=.5cm
 \begin{ruledtabular}
 \begin{tabular}{ccc}
 \hline\noalign{\smallskip}
  & BSk6 & BSk8 \\
 \noalign{\smallskip}\hline\noalign{\smallskip}
  $a_v$ {\scriptsize [MeV]} & -15.749 & -15.824   \\
  $\rho_0$ {\scriptsize [fm$^{-3}$]} & 0.1575 & 0.1589  \\
  $J$ {\scriptsize [MeV]} & 28.0 & 28.0   \\
  $M^*_s/M$ & 0.80 & 0.80   \\
  $M^*_v/M$ & 0.86 & 0.87  \\
  $K_v$ \scriptsize [MeV] & 229  & 230  \\
  $G_0$      &  0.066  & 0.042   \\
  $G_0^{'}$  & 0.312   & 0.261   \\
  $\rho_{frmg}/\rho_0$   & 1.82 & 1.70  \\
  $a_{sf}$ {\scriptsize [MeV]} & 17.18 & 17.64   \\
  $Q$ {\scriptsize [MeV]} & 53.4 & 53.9  \\
 \noalign{\smallskip}\hline
 \end{tabular}
 \end{ruledtabular}
 \end{table}
The Skyrme parameters have changed little from the original BSk6 Skyrme force; the
pairing strengths have instead decreased significantly, by 2$\%$ for $V_{\pi n}$ and 
11$\%$ for $V_{\pi p}$. For the first time we obtain $V_{\pi p} < V_{\pi n}$,
which is what one expects when the zero-range pairing force with different
coupling constants for protons and neutrons, Eq.~(\ref{2}), shall simulate 
an attractive isospin-invariant isovector pairing force and a repulsive
isospin symmetry-breaking pairing force originating from the Coulomb interaction. 
The new Skyrme parametrization globally improves the prediction of experimental 
masses, as shown in Table~\ref{tab_rms_8}. The comparison between the HFB-6 and 
HFB-8 mass tables, in particular towards the neutron dripline is further discussed 
in Sect.~\ref{sect4}.
\begin{figure}[ht]
\centerline{\epsfig{figure=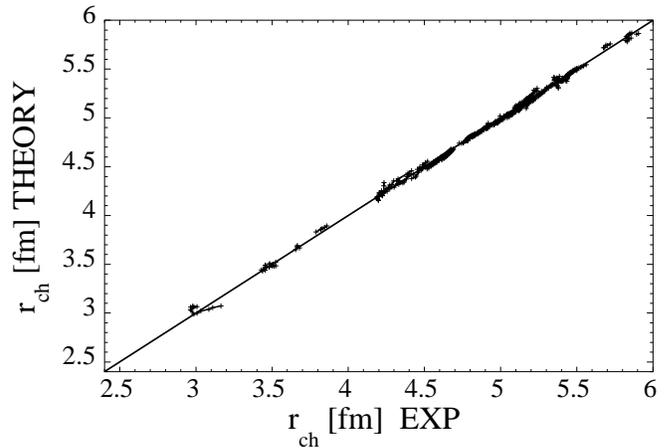,scale=.38}}
\caption{\label{fig_radius} Comparison of the theoretical 
and experimental charge radii for the 523 nuclei listed in the 1994 compilation of
\cite{nm94}} 
\end{figure}

\begin{figure}[ht]
\centerline{\epsfig{figure=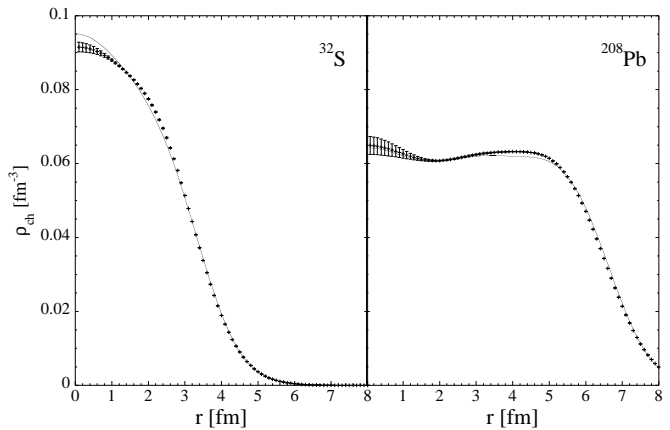,scale=.34}}
\caption{\label{fig_density} Comparison of the theoretical (full line)
and experimental (crosses with error bars) charge densities for for $^{32}$S and
$^{208}$Pb. Experimental data  are from \cite{fricke95}.} 
\end{figure}

Table~\ref{tab_rms_8} also shows the rms and mean deviations between theoretical 
and experimental charge radii for the 523 nuclei listed in the 1994 compilation
\cite{nm94} (for more details on the HFB derivation of the charge radii, see 
\cite{bp01}). 
The comparison is given in Fig.~\ref{fig_radius}. The overall agreement
with experiment is seen to be excellent, although 
some fine structure related to local anomalies is not reproduced in all details. 
Similarly, the radial charge density distribution of spherical nuclei as obtained from 
the charge form factor is found to be well reproduced even down to the centre 
of the nucleus, see the examples of $^{32}$S and $^{208}$Pb given in
Fig.~\ref{fig_density}. Such a good agreement between theoretical and experimental
densities and radii is strongly related to the adopted value of the saturation density
$\rho_0$ (Table III) or equivalently the Fermi momentum $k_F=1.33$~fm$^{-1}$ given by
$(3\pi^2\rho_0/2)^{1/3}$. However, in the case of densities, the excellent agreement with
experiment both in the surface and inside the nucleus must be regarded as fortuitous,
being completely beyond our control in mass fits of the kind described here.

\section{Extrapolations}
\label{sect4}
With the BSk8 Skyrme force determined as described  we constructed a complete mass
table, labeled HFB-8, for the same nuclei as were included in the HFB-6 tables, i.e.,
all the nuclei  lying between the two drip lines over the range of $Z$ and $N \ge 8$ and
$Z \le 120$.

\begin{figure}[t!]
\centerline{\epsfig{figure=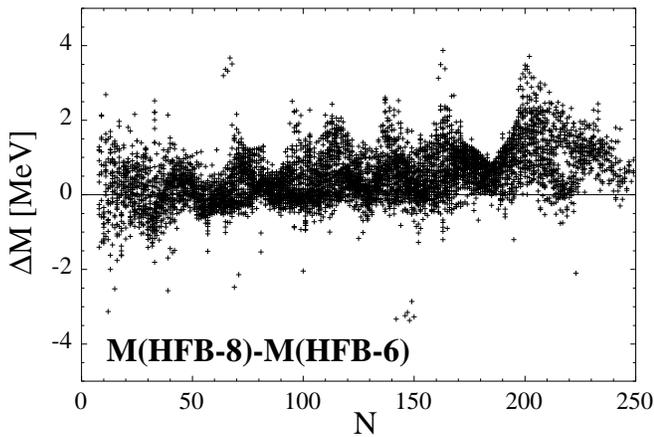,scale=.36}}
\caption{\label{fig_mdifn} Differences between the HFB-8 and the HFB-6 masses as a
function of the neutron number $N$  for all nuclei with
$8\leq Z\leq 120$  lying between the two drip lines.} 
\end{figure}

\begin{figure}[t!]
\centerline{\epsfig{figure=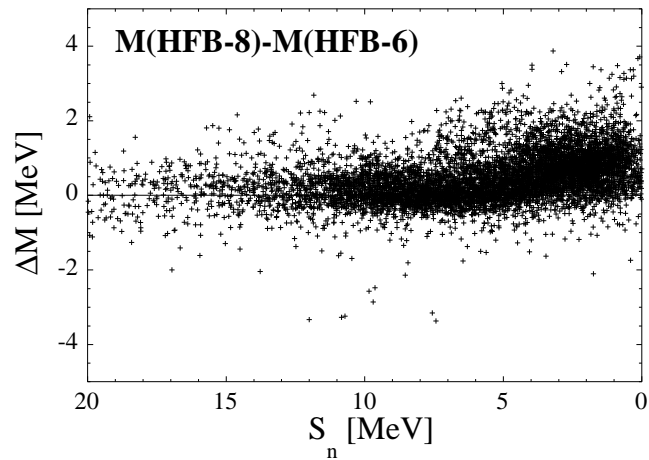,scale=.36}}
\caption{\label{fig_mdifsn} Differences between the HFB-8 and the HFB-6 masses as a
function of the neutron separation energy $S_n$ for all nuclei with
$8\leq Z\leq 120$  lying between the two drip lines.} 
\end{figure}

The HFB-8 masses are compared in Figs.~\ref{fig_mdifn}-\ref{fig_mdifsn} with 
the HFB-6 masses. 
 
Differences seldom exceed some 2~MeV, even close to the
neutron dripline; the largest deviations are found for open shell nuclei. Shell
effects far away from stability are found to be very similar for the two mass
tables. Moreover, the HFB-8 shell gaps are very similar to the shell gaps obtained with 
the HFB-6 mass formula so that they are not shown here.
These results once again confirm the relative stability of the HFB mass
predictions with respect to different parametrizations or frameworks, as already
emphasized in \cite{gs03}.

%
%********************************************************************************************
%
\section{Conclusions}
\label{concl}
The pairing correlations included in the HFB wave function are known to break the
particle-number symmetry. The restoration of the exact particle number is done on the
basis of the projection technique, i.e by projecting the wave function on the exact
number of particles after a variation that includes the approximate LN projection before
variation. Doing so, we have constructed a new Skyrme
force, labelled BSk8, the parameters of which reproduce the 2149 measured masses with an
rms error of 0.635~MeV.  The final table, referred to as HFB-8, 
includes all the 9200 nuclei lying between the two drip lines over the range of $Z$ and
$N \ge 8$ and $Z
\le 120$. The extrapolations of this new mass formula out to the drip lines do not differ
significantly from the previous HFB-6 mass formula obtained without the restoration of
the particle-number symmetry.\\
\noindent{\bf Acknowledgements.}
M.S.\ and S.G.\ are FNRS Research Fellow and Associate, respectively. 
J.M.P.\ acknowledges financial support from NSERC (Canada).
This work has been partly supported by Grant No.\ PAI-P5-07 of the 
Belgian Office for Scientific Policy.
M.B.\ acknowledges financial support from the European Community.


\begin{thebibliography}{41}
\expandafter\ifx\csname natexlab\endcsname\relax\def\natexlab#1{#1}\fi
\expandafter\ifx\csname bibnamefont\endcsname\relax
  \def\bibnamefont#1{#1}\fi
\expandafter\ifx\csname bibfnamefont\endcsname\relax
  \def\bibfnamefont#1{#1}\fi
\expandafter\ifx\csname citenamefont\endcsname\relax
  \def\citenamefont#1{#1}\fi
\expandafter\ifx\csname url\endcsname\relax
  \def\url#1{\texttt{#1}}\fi
\expandafter\ifx\csname urlprefix\endcsname\relax\def\urlprefix{URL }\fi
\providecommand{\bibinfo}[2]{#2}
\providecommand{\eprint}[2][]{\url{#2}}

\bibitem[{\citenamefont{{Samyn} et~al.}(2002)\citenamefont{{Samyn}, {Goriely},
  {Heenen}, {Pearson}, and {Tondeur}}}]{sghpt02}
\bibinfo{author}{\bibfnamefont{M.}~\bibnamefont{{Samyn}}},
  \bibinfo{author}{\bibfnamefont{S.}~\bibnamefont{{Goriely}}},
  \bibinfo{author}{\bibfnamefont{P.-H.} \bibnamefont{{Heenen}}},
  \bibinfo{author}{\bibfnamefont{J.~M.} \bibnamefont{{Pearson}}},
  \bibnamefont{and}
  \bibinfo{author}{\bibfnamefont{F.}~\bibnamefont{{Tondeur}}},
  \bibinfo{journal}{Nucl. Phys.} \textbf{\bibinfo{volume}{A700}},
  \bibinfo{pages}{142} (\bibinfo{year}{2002}).

\bibitem[{\citenamefont{{Goriely} et~al.}(2002)\citenamefont{{Goriely},
  {Samyn}, {Heenen}, {Pearson}, and {Tondeur}}}]{gshpt02}
\bibinfo{author}{\bibfnamefont{S.}~\bibnamefont{{Goriely}}},
  \bibinfo{author}{\bibfnamefont{M.}~\bibnamefont{{Samyn}}},
  \bibinfo{author}{\bibfnamefont{P.-H.} \bibnamefont{{Heenen}}},
  \bibinfo{author}{\bibfnamefont{J.~M.} \bibnamefont{{Pearson}}},
  \bibnamefont{and}
  \bibinfo{author}{\bibfnamefont{F.}~\bibnamefont{{Tondeur}}},
  \bibinfo{journal}{Phys. Rev. C} \textbf{\bibinfo{volume}{66}},
  \bibinfo{pages}{024326} (\bibinfo{year}{2002}).

\bibitem[{\citenamefont{{Samyn} et~al.}(2003)\citenamefont{{Samyn}, {Goriely},
  and {Pearson}}}]{sg03}
\bibinfo{author}{\bibfnamefont{M.}~\bibnamefont{{Samyn}}},
  \bibinfo{author}{\bibfnamefont{S.}~\bibnamefont{{Goriely}}},
  \bibnamefont{and} \bibinfo{author}{\bibfnamefont{J.~M.}
  \bibnamefont{{Pearson}}}, \bibinfo{journal}{Nucl. Phys.}
  \textbf{\bibinfo{volume}{A725}}, \bibinfo{pages}{69} (\bibinfo{year}{2003}).

\bibitem[{\citenamefont{{Goriely} et~al.}(2003)\citenamefont{{Goriely},
  {Samyn}, {Bender}, and {Pearson}}}]{gs03}
\bibinfo{author}{\bibfnamefont{S.}~\bibnamefont{{Goriely}}},
  \bibinfo{author}{\bibfnamefont{M.}~\bibnamefont{{Samyn}}},
  \bibinfo{author}{\bibfnamefont{M.}~\bibnamefont{{Bender}}}, \bibnamefont{and}
  \bibinfo{author}{\bibfnamefont{J.~M.} \bibnamefont{{Pearson}}},
  \bibinfo{journal}{Phys. Rev. C} \textbf{\bibinfo{volume}{68}},
  \bibinfo{pages}{054325} (\bibinfo{year}{2003}).

\bibitem[{\citenamefont{{Audi} and {Wapstra}}(1995)}]{aw95}
\bibinfo{author}{\bibfnamefont{G.}~\bibnamefont{{Audi}}} \bibnamefont{and}
  \bibinfo{author}{\bibfnamefont{A.}~\bibnamefont{{Wapstra}}},
  \bibinfo{journal}{Nucl. Phys.} \textbf{\bibinfo{volume}{A595}},
  \bibinfo{pages}{409} (\bibinfo{year}{1995}), \bibinfo{note}{{\it
  www-csnsm.in2p3.fr/AMDC/}}.

\bibitem[{\citenamefont{{Bulgac}}(2002)}]{b02}
\bibinfo{author}{\bibfnamefont{A.}~\bibnamefont{{Bulgac}}},
  \bibinfo{journal}{Phys. Rev. C} \textbf{\bibinfo{volume}{65}},
  \bibinfo{pages}{051305} (\bibinfo{year}{2002}).

\bibitem[{\citenamefont{{Bulgac} and {Yu}}(2002)}]{by02}
\bibinfo{author}{\bibfnamefont{A.}~\bibnamefont{{Bulgac}}} \bibnamefont{and}
  \bibinfo{author}{\bibfnamefont{Y.}~\bibnamefont{{Yu}}},
  \bibinfo{journal}{Phys. Rev. Lett.} \textbf{\bibinfo{volume}{88}},
  \bibinfo{pages}{042504} (\bibinfo{year}{2002}).

\bibitem[{\citenamefont{{Ring} and {Schuck}}(1980)}]{rs80}
\bibinfo{author}{\bibfnamefont{P.}~\bibnamefont{{Ring}}} \bibnamefont{and}
  \bibinfo{author}{\bibfnamefont{P.}~\bibnamefont{{Schuck}}},
  \emph{\bibinfo{title}{The {N}uclear {M}any-{B}ody {P}roblem}}
  (\bibinfo{publisher}{Springer, Berlin}, \bibinfo{year}{1980}).

\bibitem[{\citenamefont{{Bender} et~al.}(2003)\citenamefont{{Bender}, {Heenen},
  and {Reinhard}}}]{bh03}
\bibinfo{author}{\bibfnamefont{M.}~\bibnamefont{{Bender}}},
  \bibinfo{author}{\bibfnamefont{P.-H.} \bibnamefont{{Heenen}}},
  \bibnamefont{and} \bibinfo{author}{\bibfnamefont{P.-G.}
  \bibnamefont{{Reinhard}}}, \bibinfo{journal}{Rev. Mod. Phys.}
  \textbf{\bibinfo{volume}{75}}, \bibinfo{pages}{121} (\bibinfo{year}{2003}).

\bibitem[{\citenamefont{{Tondeur} et~al.}(2000)\citenamefont{{Tondeur},
  {Goriely}, {Pearson}, and {Onsi}}}]{tgpo00}
\bibinfo{author}{\bibfnamefont{F.}~\bibnamefont{{Tondeur}}},
  \bibinfo{author}{\bibfnamefont{S.}~\bibnamefont{{Goriely}}},
  \bibinfo{author}{\bibfnamefont{J.~M.} \bibnamefont{{Pearson}}},
  \bibnamefont{and} \bibinfo{author}{\bibfnamefont{M.}~\bibnamefont{{Onsi}}},
  \bibinfo{journal}{Phys. Rev. C} \textbf{\bibinfo{volume}{62}},
  \bibinfo{pages}{024308} (\bibinfo{year}{2000}).

\bibitem[{\citenamefont{{Butler} et~al.}(1984)\citenamefont{{Butler}, {Sprung},
  and {Martorell}}}]{bs84}
\bibinfo{author}{\bibfnamefont{M.}~\bibnamefont{{Butler}}},
  \bibinfo{author}{\bibfnamefont{D.}~\bibnamefont{{Sprung}}}, \bibnamefont{and}
  \bibinfo{author}{\bibfnamefont{J.}~\bibnamefont{{Martorell}}},
  \bibinfo{journal}{Nucl. Phys.} \textbf{\bibinfo{volume}{A422}},
  \bibinfo{pages}{157} (\bibinfo{year}{1984}).

\bibitem[{\citenamefont{{Heenen} et~al.}(1993)\citenamefont{{Heenen}, {Bonche},
  {Dobaczewski}, and {Flocard}}}]{hb93}
\bibinfo{author}{\bibfnamefont{P.-H.} \bibnamefont{{Heenen}}},
  \bibinfo{author}{\bibfnamefont{P.}~\bibnamefont{{Bonche}}},
  \bibinfo{author}{\bibfnamefont{J.}~\bibnamefont{{Dobaczewski}}},
  \bibnamefont{and}
  \bibinfo{author}{\bibfnamefont{H.}~\bibnamefont{{Flocard}}},
  \bibinfo{journal}{Nucl. Phys.} \textbf{\bibinfo{volume}{A561}},
  \bibinfo{pages}{367} (\bibinfo{year}{1993}).

\bibitem[{\citenamefont{{Anguiano} et~al.}(2001)\citenamefont{{Anguiano},
  {Egido}, and {Robledo}}}]{ae01}
\bibinfo{author}{\bibfnamefont{M.}~\bibnamefont{{Anguiano}}},
  \bibinfo{author}{\bibfnamefont{J.}~\bibnamefont{{Egido}}}, \bibnamefont{and}
  \bibinfo{author}{\bibfnamefont{L.}~\bibnamefont{{Robledo}}},
  \bibinfo{journal}{Nucl. Phys.} \textbf{\bibinfo{volume}{A696}},
  \bibinfo{pages}{467} (\bibinfo{year}{2001}).

\bibitem[{\citenamefont{{Anguiano} et~al.}(2002)\citenamefont{{Anguiano},
  {Egido}, and {Robledo}}}]{ae02}
\bibinfo{author}{\bibfnamefont{M.}~\bibnamefont{{Anguiano}}},
  \bibinfo{author}{\bibfnamefont{J.}~\bibnamefont{{Egido}}}, \bibnamefont{and}
  \bibinfo{author}{\bibfnamefont{L.}~\bibnamefont{{Robledo}}},
  \bibinfo{journal}{Phys. Lett. B} \textbf{\bibinfo{volume}{545}},
  \bibinfo{pages}{62} (\bibinfo{year}{2002}).

\bibitem[{\citenamefont{Fomenko}(1970)}]{fo70}
\bibinfo{author}{\bibfnamefont{V.~N.} \bibnamefont{Fomenko}},
  \bibinfo{journal}{J. Phys. (London)} \textbf{\bibinfo{volume}{A3}},
  \bibinfo{pages}{8} (\bibinfo{year}{1970}).

\bibitem[{\citenamefont{{Nogami}}(1973)}]{N64}
\bibinfo{author}{\bibfnamefont{Y.}~\bibnamefont{{Nogami}}},
  \bibinfo{journal}{Phys. Rev.} \textbf{\bibinfo{volume}{134}},
  \bibinfo{pages}{B313} (\bibinfo{year}{1973}).

\bibitem[{\citenamefont{{Pradhan} et~al.}(1973)\citenamefont{{Pradhan},
  {Nogami}, and {Law}}}]{PN73}
\bibinfo{author}{\bibfnamefont{H.}~\bibnamefont{{Pradhan}}},
  \bibinfo{author}{\bibfnamefont{Y.}~\bibnamefont{{Nogami}}}, \bibnamefont{and}
  \bibinfo{author}{\bibfnamefont{J.}~\bibnamefont{{Law}}},
  \bibinfo{journal}{nucl. Phys. A} \textbf{\bibinfo{volume}{201}},
  \bibinfo{pages}{357} (\bibinfo{year}{1973}).

\bibitem[{\citenamefont{Reinhard et~al.}(1996)\citenamefont{Reinhard,
  Nazarewicz, Bender, and Maruhn}}]{rn96}
\bibinfo{author}{\bibfnamefont{P.-G.} \bibnamefont{Reinhard}},
  \bibinfo{author}{\bibfnamefont{W.}~\bibnamefont{Nazarewicz}},
  \bibinfo{author}{\bibfnamefont{M.}~\bibnamefont{Bender}}, \bibnamefont{and}
  \bibinfo{author}{\bibfnamefont{J.~A.} \bibnamefont{Maruhn}},
  \bibinfo{journal}{Phys. Rev. C} \textbf{\bibinfo{volume}{53}},
  \bibinfo{pages}{2776} (\bibinfo{year}{1996}).

\bibitem[{\citenamefont{Flocard and Onishi}(1997)}]{fo97}
\bibinfo{author}{\bibfnamefont{H.}~\bibnamefont{Flocard}} \bibnamefont{and}
  \bibinfo{author}{\bibfnamefont{N.}~\bibnamefont{Onishi}},
  \bibinfo{journal}{Ann. Phys.} \textbf{\bibinfo{volume}{254}},
  \bibinfo{pages}{275} (\bibinfo{year}{1997}).

\bibitem[{\citenamefont{Rodriguez-Guzman
  et~al.}(2002)\citenamefont{Rodriguez-Guzman, Egido, and Robledo}}]{ro02}
\bibinfo{author}{\bibfnamefont{R.}~\bibnamefont{Rodriguez-Guzman}},
  \bibinfo{author}{\bibfnamefont{J.~L.} \bibnamefont{Egido}}, \bibnamefont{and}
  \bibinfo{author}{\bibfnamefont{L.~M.} \bibnamefont{Robledo}},
  \bibinfo{journal}{Nucl. Phys.} \textbf{\bibinfo{volume}{A709}},
  \bibinfo{pages}{201} (\bibinfo{year}{2002}).

\bibitem[{\citenamefont{{Inglis}}(1954)}]{in54}
\bibinfo{author}{\bibfnamefont{D.}~\bibnamefont{{Inglis}}},
  \bibinfo{journal}{Phys. Rev.} \textbf{\bibinfo{volume}{96}},
  \bibinfo{pages}{1059} (\bibinfo{year}{1954}).

\bibitem[{\citenamefont{{Inglis}}(1956)}]{in56}
\bibinfo{author}{\bibfnamefont{D.}~\bibnamefont{{Inglis}}},
  \bibinfo{journal}{Phys. Rev.} \textbf{\bibinfo{volume}{103}},
  \bibinfo{pages}{1786} (\bibinfo{year}{1956}).

\bibitem[{\citenamefont{Belyaev}(1961)}]{be61}
\bibinfo{author}{\bibfnamefont{S.}~\bibnamefont{Belyaev}},
  \bibinfo{journal}{Nucl. Phys.} \textbf{\bibinfo{volume}{24}},
  \bibinfo{pages}{322} (\bibinfo{year}{1961}).

\bibitem[{\citenamefont{{Pearson} et~al.}(1991)\citenamefont{{Pearson},
  {Aboussir}, {Dutta}, {Nayak}, {Farine}, and {Tondeur}}}]{pa91}
\bibinfo{author}{\bibfnamefont{J.~M.} \bibnamefont{{Pearson}}},
  \bibinfo{author}{\bibfnamefont{Y.}~\bibnamefont{{Aboussir}}},
  \bibinfo{author}{\bibfnamefont{A.~K.} \bibnamefont{{Dutta}}},
  \bibinfo{author}{\bibfnamefont{R.~C.} \bibnamefont{{Nayak}}},
  \bibinfo{author}{\bibfnamefont{M.}~\bibnamefont{{Farine}}}, \bibnamefont{and}
  \bibinfo{author}{\bibfnamefont{F.}~\bibnamefont{{Tondeur}}},
  \bibinfo{journal}{Nucl. Phys.} \textbf{\bibinfo{volume}{A528}},
  \bibinfo{pages}{1} (\bibinfo{year}{1991}).

\bibitem[{\citenamefont{{Goriely} et~al.}(2001)\citenamefont{{Goriely},
  {Tondeur}, and {Pearson}}}]{gtp01}
\bibinfo{author}{\bibfnamefont{S.}~\bibnamefont{{Goriely}}},
  \bibinfo{author}{\bibfnamefont{F.}~\bibnamefont{{Tondeur}}},
  \bibnamefont{and} \bibinfo{author}{\bibfnamefont{J.~M.}
  \bibnamefont{{Pearson}}}, \bibinfo{journal}{At. Data Nucl. Data Tables}
  \textbf{\bibinfo{volume}{77}}, \bibinfo{pages}{311} (\bibinfo{year}{2001}).

\bibitem[{\citenamefont{{Bj\o rnholm} and {Lynn}}(1980)}]{bl80}
\bibinfo{author}{\bibfnamefont{S.}~\bibnamefont{{Bj\o rnholm}}}
  \bibnamefont{and} \bibinfo{author}{\bibfnamefont{J.}~\bibnamefont{{Lynn}}},
  \bibinfo{journal}{Rev. Mod. Phys.} \textbf{\bibinfo{volume}{52}},
  \bibinfo{pages}{725} (\bibinfo{year}{1980}).

\bibitem[{\citenamefont{Blons et~al.}(1988)\citenamefont{Blons, Fabbro, Mazur,
  Paya, and Ribrag}}]{bf88}
\bibinfo{author}{\bibfnamefont{J.}~\bibnamefont{Blons}},
  \bibinfo{author}{\bibfnamefont{B.}~\bibnamefont{Fabbro}},
  \bibinfo{author}{\bibfnamefont{C.}~\bibnamefont{Mazur}},
  \bibinfo{author}{\bibfnamefont{D.}~\bibnamefont{Paya}}, \bibnamefont{and}
  \bibinfo{author}{\bibfnamefont{M.}~\bibnamefont{Ribrag}},
  \bibinfo{journal}{Nucl. Phys.} \textbf{\bibinfo{volume}{A477}},
  \bibinfo{pages}{231} (\bibinfo{year}{1988}).

\bibitem[{jae(1996)}]{jaeri}
\emph{\bibinfo{title}{Experimental total half-lifes}},
  \bibinfo{organization}{Japan Atomic Energy Research Institute}
  (\bibinfo{year}{1996}).

\bibitem[{\citenamefont{{Hunyadi} et~al.}(2001)\citenamefont{{Hunyadi},
  {Gassmann}, {Krasznahorkay}, {Habs}, {Thirolf}, {Csatl\'os}, {Eisermann},
  {Faestermann}, {Graw}, {Guly\'as} et~al.}}]{hg01}
\bibinfo{author}{\bibfnamefont{M.}~\bibnamefont{{Hunyadi}}},
  \bibinfo{author}{\bibfnamefont{D.}~\bibnamefont{{Gassmann}}},
  \bibinfo{author}{\bibfnamefont{A.}~\bibnamefont{{Krasznahorkay}}},
  \bibinfo{author}{\bibfnamefont{D.}~\bibnamefont{{Habs}}},
  \bibinfo{author}{\bibfnamefont{P.}~\bibnamefont{{Thirolf}}},
  \bibinfo{author}{\bibfnamefont{M.}~\bibnamefont{{Csatl\'os}}},
  \bibinfo{author}{\bibfnamefont{Y.}~\bibnamefont{{Eisermann}}},
  \bibinfo{author}{\bibfnamefont{T.}~\bibnamefont{{Faestermann}}},
  \bibinfo{author}{\bibfnamefont{G.}~\bibnamefont{{Graw}}},
  \bibinfo{author}{\bibfnamefont{J.}~\bibnamefont{{Guly\'as}}},
  \bibnamefont{et~al.}, \bibinfo{journal}{Phys. Lett. B}
  \textbf{\bibinfo{volume}{505}}, \bibinfo{pages}{27} (\bibinfo{year}{2001}).

\bibitem[{\citenamefont{Stetcu and Johnson}(2002)}]{st02}
\bibinfo{author}{\bibfnamefont{I.}~\bibnamefont{Stetcu}} \bibnamefont{and}
  \bibinfo{author}{\bibfnamefont{C.~W.} \bibnamefont{Johnson}},
  \bibinfo{journal}{Phys. Rev. C} \textbf{\bibinfo{volume}{66}},
  \bibinfo{pages}{034301} (\bibinfo{year}{2002}).

\bibitem[{\citenamefont{Baroni et~al.}()\citenamefont{Baroni, Armati, Barranco,
  Broglia, Colo', Gori, and Vigezzi}}]{ba04}
\bibinfo{author}{\bibfnamefont{S.}~\bibnamefont{Baroni}},
  \bibinfo{author}{\bibfnamefont{M.}~\bibnamefont{Armati}},
  \bibinfo{author}{\bibfnamefont{F.}~\bibnamefont{Barranco}},
  \bibinfo{author}{\bibfnamefont{R.~A.} \bibnamefont{Broglia}},
  \bibinfo{author}{\bibfnamefont{G.}~\bibnamefont{Colo'}},
  \bibinfo{author}{\bibfnamefont{G.}~\bibnamefont{Gori}}, \bibnamefont{and}
  \bibinfo{author}{\bibfnamefont{E.}~\bibnamefont{Vigezzi}},
  \bibinfo{note}{nucl-th/0404019}.

\bibitem[{\citenamefont{Bender et~al.}(2004)\citenamefont{Bender, Bertsch, and
  Heenen}}]{bbh03}
\bibinfo{author}{\bibfnamefont{M.}~\bibnamefont{Bender}},
  \bibinfo{author}{\bibfnamefont{G.}~\bibnamefont{Bertsch}}, \bibnamefont{and}
  \bibinfo{author}{\bibfnamefont{P.-H.} \bibnamefont{Heenen}},
  \bibinfo{journal}{Phys. Rev. C} \textbf{\bibinfo{volume}{69}},
  \bibinfo{pages}{034340} (\bibinfo{year}{2004}).

\bibitem[{\citenamefont{{Audi} and {Wapstra}}(2001)}]{aw01}
\bibinfo{author}{\bibfnamefont{G.}~\bibnamefont{{Audi}}} \bibnamefont{and}
  \bibinfo{author}{\bibfnamefont{A.}~\bibnamefont{{Wapstra}}}
  (\bibinfo{year}{2001}), \bibinfo{note}{private communication}.

\bibitem[{\citenamefont{{Audi} et~al.}(2003)\citenamefont{{Audi}, {Wapstra},
  and {Thibault}}}]{aw03}
\bibinfo{author}{\bibfnamefont{G.}~\bibnamefont{{Audi}}},
  \bibinfo{author}{\bibfnamefont{A.}~\bibnamefont{{Wapstra}}},
  \bibnamefont{and}
  \bibinfo{author}{\bibfnamefont{C.}~\bibnamefont{{Thibault}}},
  \bibinfo{journal}{Nucl. Phys.} \textbf{\bibinfo{volume}{A729}},
  \bibinfo{pages}{3} (\bibinfo{year}{2003}), \bibinfo{note}{{\it
  www-csnsm.in2p3.fr/AMDC}}.

\bibitem[{\citenamefont{{Nguyen Van Giai} and {Sagawa}}(1981)}]{gs81}
\bibinfo{author}{\bibnamefont{{Nguyen Van Giai}}} \bibnamefont{and}
  \bibinfo{author}{\bibfnamefont{H.}~\bibnamefont{{Sagawa}}},
  \bibinfo{journal}{Phys. Lett. B} \textbf{\bibinfo{volume}{106}},
  \bibinfo{pages}{379} (\bibinfo{year}{1981}).

\bibitem[{\citenamefont{{Kutschera} and {W\'ojcik}}(1994)}]{kw94}
\bibinfo{author}{\bibfnamefont{M.}~\bibnamefont{{Kutschera}}} \bibnamefont{and}
  \bibinfo{author}{\bibfnamefont{W.}~\bibnamefont{{W\'ojcik}}},
  \bibinfo{journal}{Phys. Lett. B} \textbf{\bibinfo{volume}{325}},
  \bibinfo{pages}{271} (\bibinfo{year}{1994}).

\bibitem[{\citenamefont{Myers and Swiatecki}(1969)}]{ms69}
\bibinfo{author}{\bibfnamefont{W.~D.} \bibnamefont{Myers}} \bibnamefont{and}
  \bibinfo{author}{\bibfnamefont{W.~J.} \bibnamefont{Swiatecki}},
  \bibinfo{journal}{Ann. Phys.} \textbf{\bibinfo{volume}{55}},
  \bibinfo{pages}{395} (\bibinfo{year}{1969}).

\bibitem[{\citenamefont{Bender et~al.}(2002)\citenamefont{Bender, Dobaczewski,
  Engel, and Nazarewicz}}]{bd02}
\bibinfo{author}{\bibfnamefont{M.}~\bibnamefont{Bender}},
  \bibinfo{author}{\bibfnamefont{J.}~\bibnamefont{Dobaczewski}},
  \bibinfo{author}{\bibfnamefont{J.}~\bibnamefont{Engel}}, \bibnamefont{and}
  \bibinfo{author}{\bibfnamefont{W.}~\bibnamefont{Nazarewicz}},
  \bibinfo{journal}{Phys. Rev. C} \textbf{\bibinfo{volume}{65}},
  \bibinfo{pages}{054322} (\bibinfo{year}{2002}).

\bibitem[{\citenamefont{{Nadjakov} et~al.}(1994)\citenamefont{{Nadjakov},
  {Marinova}, and {Gangrsky}}}]{nm94}
\bibinfo{author}{\bibfnamefont{E.}~\bibnamefont{{Nadjakov}}},
  \bibinfo{author}{\bibfnamefont{K.}~\bibnamefont{{Marinova}}},
  \bibnamefont{and}
  \bibinfo{author}{\bibfnamefont{Y.}~\bibnamefont{{Gangrsky}}},
  \bibinfo{journal}{At Data Nucl. Data Tables} \textbf{\bibinfo{volume}{56}},
  \bibinfo{pages}{133} (\bibinfo{year}{1994}).

\bibitem[{\citenamefont{{Fricke} et~al.}(1995)\citenamefont{{Fricke},
  {Bernhardt}, {Heilig}, {Schaller}, {Shellenberger}, {Shera}, and {de
  Jager}}}]{fricke95}
\bibinfo{author}{\bibfnamefont{G.}~\bibnamefont{{Fricke}}},
  \bibinfo{author}{\bibfnamefont{C.}~\bibnamefont{{Bernhardt}}},
  \bibinfo{author}{\bibfnamefont{K.}~\bibnamefont{{Heilig}}},
  \bibinfo{author}{\bibfnamefont{L.}~\bibnamefont{{Schaller}}},
  \bibinfo{author}{\bibfnamefont{L.}~\bibnamefont{{Shellenberger}}},
  \bibinfo{author}{\bibfnamefont{E.}~\bibnamefont{{Shera}}}, \bibnamefont{and}
  \bibinfo{author}{\bibfnamefont{C.}~\bibnamefont{{de Jager}}},
  \bibinfo{journal}{At. Data Nucl. Data Tables} \textbf{\bibinfo{volume}{60}},
  \bibinfo{pages}{177} (\bibinfo{year}{1995}).

\bibitem[{\citenamefont{Buchinger et~al.}(2001)\citenamefont{Buchinger,
  Pearson, and Goriely}}]{bp01}
\bibinfo{author}{\bibfnamefont{F.}~\bibnamefont{Buchinger}},
  \bibinfo{author}{\bibfnamefont{J.~M.} \bibnamefont{Pearson}},
  \bibnamefont{and} \bibinfo{author}{\bibfnamefont{S.}~\bibnamefont{Goriely}},
  \bibinfo{journal}{Phys. Rev. C} \textbf{\bibinfo{volume}{64}},
  \bibinfo{pages}{067303} (\bibinfo{year}{2001}).

\end{thebibliography}
\end{document}